\documentclass[aps,prd,nofootinbib,preprint,superscriptaddress]{revtex4}
\usepackage[utf8]{inputenc}
\usepackage[english]{babel}
\usepackage[fleqn]{amsmath}
\usepackage{amsfonts}
\usepackage{amssymb}
\usepackage[section]{placeins}
\usepackage{siunitx}
\usepackage{xifthen}
\usepackage{multirow,listliketab,tabu}
\usepackage{array}
\usepackage[expansion=true]{microtype}	
\usepackage{mathtools,colonequals,bm,rotating}
\usepackage{etextools}
\usepackage{tikz}
\usetikzlibrary{decorations}
\usetikzlibrary{decorations.pathmorphing}
\usepackage{tkz-euclide}

\usepackage{graphicx,picture,placeins}
\usepackage{multirow}
\usepackage[hypertexnames=false]{hyperref}
\hypersetup{
  colorlinks = true,
  urlcolor   = black,
  linkcolor  = black,
  citecolor  = black
}
\graphicspath{{./Plots/}}
\newcommand*{\eh}[1]{\mathrm e^{#1}}

\newcommand*{\diff}{\mathrm d}

\newcommand*{\lint}{\int\limits}

\DeclareMathOperator*{\arsinh}{arsinh}
\newcommand{\eperp}{{\epsilon_\perp}}

\renewcommand{\vec}[1]{{\bm #1}}
\renewcommand{\Re}{\operatorname{Re}}
\renewcommand{\Im}{\operatorname{Im}}

\newcommand{\ftt}{t_\text{f.t.}}
\newcommand{\env}{\mathcal E}
\allowdisplaybreaks

\begin{document}

\title{Dynamical Schwinger process in a bifrequent electric field of finite
duration: survey on amplification}
\author{A.~Otto}
\affiliation{Institute of Radiation Physics, Helmholtz-Zentrum
Dresden-Rossendorf,\\Bautzner Landstra\ss e 400, 01328 Dresden, Germany}
\affiliation{Institut f\"ur Theoretische Physik, Technische Universit\"at
Dresden,\\Zellescher Weg 17, 01062 Dresden, Germany}
\author{D.~Seipt}
\affiliation{Helmholtz-Institut Jena, Fr\"obelstieg 3, 07743 Jena, Germany}
\author{D.~Blaschke}
\affiliation{Institute for Theoretical Physics, University of Wroclaw,\\
pl.\ M.\ Borna 9, 50-204 Wroclaw, Poland}
\affiliation{Bogoliubov Laboratory for Theoretical Physics, JINR Dubna,\\
Joliot-Curie str.\ 6, 141980 Dubna, Russia}
\author{S.A.~Smolyansky}
\affiliation{Department of Physics, Saratov State University, 410071 Saratov,
Russia}
\author{B.~Kämpfer}
\affiliation{Institute of Radiation Physics, Helmholtz-Zentrum
Dresden-Rossendorf,\\Bautzner Landstra\ss e 400, 01328 Dresden, Germany}
\affiliation{Institut f\"ur Theoretische Physik, Technische Universit\"at
Dresden,\\Zellescher Weg 17, 01062 Dresden, Germany}

\date{\today}

\begin{abstract}
The electron-positron pair production due to the dynamical Schwinger process in
a slowly oscillating strong electric field is enhanced by the superposition of a
rapidly oscillating weaker electric field.
A systematic account of the enhancement by the resulting bifrequent field is
provided for the residual phase space distribution.
The enhancement is explained by a severe reduction of the suppression in both
the tunneling and multiphoton regimes.
\end{abstract}

\maketitle

\section{Introduction}
The developement of new radiation sources offers opportunities for the
investigation of fundamental physical processes which hitherto could not be
accessed experimentally.
In the realm of quantum electrodynamics the Schwinger effect, i.e.\ the decay of
a pure electromagnetic field into electron-positron pairs, is among the
challenges which have escaped an experimental verification until now.
While pair production via perturbative or even non-linear effects in different
forms, mostly including null fields, is well established, the genuinely
non-perturbative Schwinger effect~\cite{schwinger} in a static homogenous
electric field requires field strengths far beyond current laboratory
capabilities.
Much hope has been put on optical laser facilities where, in fact, high field
strengths can be achieved in the focal spots near to the refraction limit.
However, only future developements as, e.g., the pillar-3 part of ELI~\cite
{ELI}, could bring the dynamical Schwinger effect (which refers to an
alternating electric field) into realistic reach.
Here, synchronized counterpropagating laser beams with suitable polarisation
can produce regions of an alternating field with a dominating electric component
in the vicinity of the antinodes.

In the search for configurations which could enable a verification of the
Schwinger effect as a tunneling process the idea has been put forward to enhance
the pair creation rate by the assistance of a multi-photon process~\cite
{schutzhold_dynamically_2008, dunne_catalysis_2009}.
This set-up is denoted as the assisted dynamical Schwinger process.
There, one can think of the combination of optical high-intensity laser and XFEL
beams, the latter ones representing a fast weak field.
In fact, at LCLS such instrumental prepositions are already at our
disposal~\cite{LCLS}, and the HIBEF collaboration~\cite{HIBEF} at the European
XFEL in Hamburg plans an analogous installation, albeit with different key
parameters w.r.t.\ repetition rates, energies and intensities.
Further experimental proposals can be found in~\cite{alkofer_pair_2001,
dunne_extreme_2014,gonoskov_probing_2013}.
Whether such configurations will enable to investigate the Schwinger effect, or
a variant thereof, needs to be elucidated.

Given such a motivation, we are going to consider here a model for pair
production by a bifrequent, spatially homogenous electric field which acts for a
finite time interval.
The pair creation process is a non-equilibrium process to be described by a
quantum kinetic equation with a strong non-Markovian feature.
Such a framework has been employed in various previous analyses, e.g., for the
superposition of Sauter pulses~\cite{orthaber_momentum_2011,
kohlfurst_optimizing_2013} or periodic fields and Sauter pulses~\cite
{sicking_bachelor_2012} or for two (or more) periodic fields~\cite
{hebenstreit_optimization_2014, akal_electron-positron_2014}, and for studying
the temporal structure of particle creation~\cite{blaschke_properties_2013,
blaschke_dynamical_2014,dabrowski_super-adiabatic_2014}.
Alternative frameworks make use of WKB-type approximations~\cite
{fey_momentum_2012}, worldline instantons~\cite{dumlu_complex_2011,
schneider_dynamically_2014} or lightfront methods~\cite{hebenstreit_pair_2011,
ilderton_localisation_2014}.
Our goal is to investigate the dynamical Schwinger effect in bifrequent fields
systematically over a large region in the parameter space spanned by field
strengths and frequencies.

\section{Residual phase space distribution: Analytical Approximations}
\begin{figure}
\centering
\includegraphics[width=\textwidth]{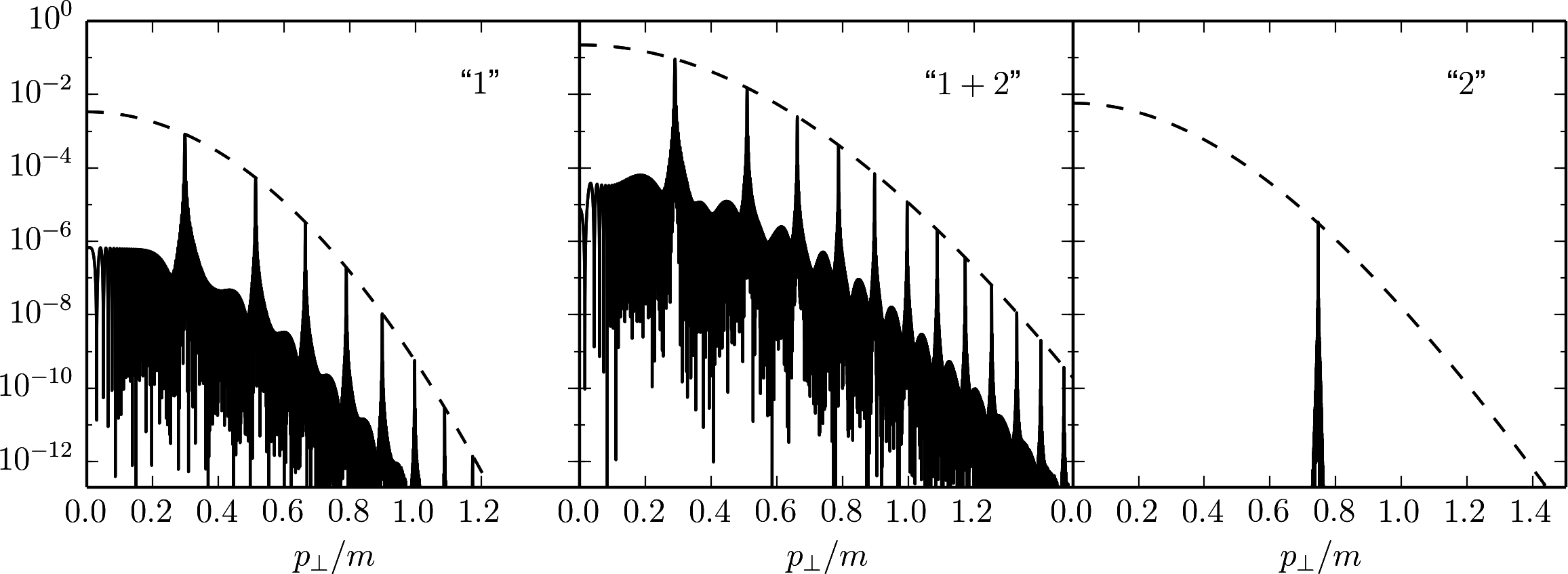}
\caption{
Residual phase space distributions at $p_\parallel=0$ (solid curves, from
integrating the quantum kinetic equations, e.g.\ equations (1-7) in~\cite
{otto_lifting_2015}) for field ``1'' (left panel, $E_1=\num{0.2}E_c$,
$\nu_1=\num{0.02}m$, i.e.\ $\gamma_1=\num{0.1}$), fields ``1+2'' (middle panel)
and field ``2'' (right panel, $E_2=\num{0.05}E_c$, $\nu_2=\num{0.5}m$,
i.e.\ $\gamma_2=\num{10}$).
The dashed curves depict the spectral envelope $\frac{2}{9}\nu_1^2\ftt^2\env(
p_\perp)$ with $\env$ given by~\eqref{eq:env}.
The flat-top duration is determined as in~\cite{otto_lifting_2015} by $\nu_1\ftt
=2\pi\cdot50$ and the $C^\infty$ ramping (switching-on) and deramping
(switching-off) periods are such to cover $5$ oscillations each.
}
\label{fig:cut_env}
\end{figure}

The residual phase space distribution $f$ of pairs in a bifrequent electric
$\vec E(t)$ field (frequencies $\nu_1$, $\nu_2=N\nu_1$) which acted for a finite
time $\ftt$ with constant amplitudes $E_1$ and $E_2$ is given by (see Appendix)
\begin{align}
f(p,\varphi,\ftt) \cong \frac{1}{2}\!\!\sum_{\ell\ge\ell_\text{min}}^\infty\!\!
\left| F_\ell\bigl(p_\ell(\varphi),\varphi\bigr) \right|^2\ftt^2
S\Bigl( \Omega'\bigl(p_\ell(\varphi),\varphi\bigr)
\ftt\bigl(p-p_\ell(\varphi)\bigr) \Bigr)\:,
\label{eq:f_shells}
\end{align}
with momentum components perpendicular ($p_\perp=p\cos\varphi$) and parallel
($p_\parallel=p\sin\varphi$) to $\vec E$ ($\varphi=-\frac{\pi}{2}\cdots
\frac{\pi}{2}$).
The function $S(x)=\sin^2x/x^2$ has a main peak at $x=0$, $S(0)=1$, of width
$2\pi$ and an infinite number of side peaks declining as $\sim x^{-2}$.
This implies that the phase space distribution~\eqref{eq:f_shells} for fixed
$\varphi$ displays a series of main peaks at $p=p_\ell(\varphi)$, see
Fig.~\ref{fig:cut_env}.
(Accounting for the $\varphi$ degree of freedom gives rise to ridges over the
$p_\perp$-$p_\parallel$ plane, sometimes called shell structure.)
The peak positions $p_\ell(\varphi)$ are determined by the resonance condition
(cf.~\cite{otto_lifting_2015})
\begin{align}
&2\Omega(p_\ell(\varphi),\varphi)-\ell\nu_1 = 0\:,\label{eq:shell_cond}\\
&\Omega(p_\perp,p_\parallel) = \frac{m}{2\pi}\lint_0^{2\pi}\!\!\diff x \left(
1+\frac{p_\perp^2}{m^2}+\left(\frac{p_\parallel}{m}-\gamma_1^{-1}\cos x
-\gamma_2^{-1}\cos Nx\right)^2 \right)^\frac{1}{2}\!,\label{eq:Omega}
\end{align}
where $\gamma_{1,2}=\frac{E_c}{E_{1,2}}\frac{\nu_{1,2}}{m}$ are Keldysh
parameters with $E_c=\frac{e}{m^2}$, $-e$ the electron charge and $m$ its mass.
Since $\Omega(p_\ell(\varphi), \varphi)>m_*>m$ with the effective mass $m_*=
\Omega(p_\ell=0)$ (cf.~\cite{brezin_pair_1970,kohlfurst_effective_2014} for the
effective mass concept in the single field case), the values of $\ell$ must
exceed a certain threshold, $\ell\ge\ell_\text{min}=\lceil\frac{2m_*}{\nu_1}
\rceil$, where $\lceil x\rceil$ denotes the smallest integer greater than or
equal to $x$.

The number of side peaks belonging to $p_\ell$ until the next main peak at
$p_{\ell+1}$ is about $k=(p_{\ell+1}-p_\ell)\,\Omega'(p_\ell(\varphi),\varphi)\,
\ftt/\pi\gg1$.
This means that the spectrum consists, for sufficiently large $\ftt$, of
well separated main peaks centered at $p_\ell(\varphi)$, with some
micro-structures emerging from the superposition of the side peaks encoded in
$S$, which are not resolved on the scale displayed in Fig.~\ref{fig:cut_env}.
Due to $\lim\limits_{\ftt\rightarrow\infty}S(a\ftt)=2\pi\ftt^{-1}\delta(a)$,
Eq.~\eqref{eq:f_shells} asymptotically gets a form derived in~\cite
{popov_resonant_1973} (see also~\cite{ringwald_pair_2001}), albeit for a single
field and in a low-momentum expansion and for the rate, which is time ($\ftt$)
independent.
We consider here finite values of $\ftt$, (i) on the one hand sufficiently large
to accomodate at least a few oscillations of both fields ``1'' and ``2'' within
the flat-top ($\ftt$) period of the overall shape function of the potential, and
(ii) not too large to avoid the onset of Rabi oscillations~\cite
{akal_electron-positron_2014} which would modify the $\ftt$ dependence in~\eqref
{eq:f_shells}.

The main peak altitudes are governed by the squared Fourier-coefficients
$\left|F_\ell\bigl(p_\ell(\varphi),\varphi\bigr)\right|^2$ which contain the
spectral envelope function $\env(p_\perp)$.
As pointed out in the Appendix, for $p_\parallel=0$, they can be approximated by
\begin{align}
\left|F_\ell(p_\ell,\varphi=0)\right|^2 &\approx \frac{2}{9}\nu_1^2\env(p_\ell)
\bigl(1-(-1)^\ell\bigr)\:,\label{eq:F_exp}\\
\env(p_\perp) &= \eh{-4\frac{m}{\nu_1}G(p_\perp,\gamma_1,\gamma_2,N)}\:,
\label{eq:env}\\
G(p_\perp,\gamma_1,\gamma_2,N) &= \lint_0^{x_0}\!\!\diff x\left( 1+\frac
{p_\perp^2}{m^2}-\left(\gamma_1^{-1}\sinh x+\gamma_2^{-1}\sinh Nx\right)^2
\right)^\frac{1}{2}\:,\label{eq:G}
\end{align}
for the special case $N=4n+1$, $n=0,1,2,\cdots$.
The quantity $x_0$ is determined by
\begin{align}
\gamma_1^{-1}\sinh x_0 + \gamma_2^{-1}\sinh Nx_0 = \sqrt{1 + p_\perp^2/m^2}\:.
\label{eq:x0}
\end{align}
In the single field case ($E_2\rightarrow0$, $\gamma_2\rightarrow\infty$), $G$
reduces to
\begin{align}
G_1(p_\perp, \gamma_1) = G(\gamma_2\rightarrow\infty) = \frac{\pi}{4}
\gamma_1 \left(1 + \frac{p_\perp^2}{m^2}\right)
g\left(\gamma_1\sqrt{1 + \frac{p_\perp^2}{m^2}}\right)\:,
\end{align}
where $g(\gamma)={_1F_2}(\frac{1}{2},\frac{1}{2},2,-\gamma^2)$ is a function
already introduced in~\cite{brezin_pair_1970}, cf.\ also~\cite
{popov_resonant_1973}.
We argue that a handy approximation is provided by
\begin{align}
G \approx \frac{\pi}{4} \sqrt{1 + \frac{p_\perp^2}{m^2}}\, x_0
\label{eq:G_approx}
\end{align}
and, analogously,
\begin{align}
G_1 \approx \frac{\pi}{4}
\sqrt{1 + \frac{p_\perp^2}{m^2}}
\arsinh\left(\sqrt{1 + \frac{p_\perp^2}{m^2}}\,\gamma_1\right)\:.
\label{eq:g_approx}
\end{align}
The former is appropriate for $x_0<1$ and the latter for
$\sqrt{1 + \frac{p_\perp^2}{m^2}}\,\gamma_1 < 10$.
The spectral envelopes are displayed in Fig.~\ref{fig:cut_env} by dashed curves
using~\eqref{eq:G}.
Evidently, the envelopes connect the peaks of the fairly rich spectra very well,
the details of which are only accessible by integrating the full set of quantum
kinetic equations.
The approximations~\eqref{eq:G_approx} and~\eqref{eq:g_approx} are suitable for
conditions as in the left panel of Fig.~\ref{fig:cut_env}, while~\eqref
{eq:G_approx} gives some semi-quantitative account (within factor of two) for
the middle panel.

To understand qualitatively the amplification by a second weak field, as
exemplified in Fig.~\ref{fig:cut_env} (further examples in~\cite
{otto_lifting_2015}), one can use the approximation~\eqref{eq:G_approx} for
$p_\perp\ll m$ and the first-order iterative solution of~\eqref{eq:x0}, $x_0=
\arsinh\left(\gamma_1-\frac{\gamma_1}{\gamma_2}\sinh\bigl[N\arsinh\gamma_1\bigr]
\right)$.
Supposing $\gamma_1\ll1$ and $N\gamma_1\ll1$ one finds for large $\gamma_2$
\begin{align}
G(\gamma_1,\gamma_2,N)\approx G_1(\gamma_1)-\frac{\gamma_1^2}{\gamma_2}N,
\end{align}
i.e.\ due to the presence of the second field, encoded in $\gamma_2$, the
modulus of the exponential, $4\frac{m}{\nu_1}G_1$, becomes diminished by about
$4\left(\frac{E_c}{E_1}\right)^2\frac{E_2}{E_c}$ -- interestingly independent of
$\nu_2$ in the given approximation (which is a special case of the expansion of
a multi-scale implicit function).
In other words, the spectral envelope of field ``1'' gets lifted by $\exp\left(4
\frac{m}{\nu_1}\frac{\gamma_1^2}{\gamma_2}N\right)$ due to the presence of
field ``2''.
In general we stress that, due to the monotonic behavior of the hyperbolic sine
functions in~\eqref{eq:G} and the defining equation~\eqref{eq:x0} for $x_0$, one
infers $x_0(\gamma_1,\gamma_2\rightarrow\infty,N)>x_0(\gamma_1,\gamma_2,N)>0$
and analogously $G_1(\gamma_1)>G(\gamma_1,\gamma_2,N)>0$.
Since the negative of $G$ enters the exponential of the envelope function, a
dropping of $G$ by the second weak field causes the anticipated amplification.
A quantitative consideration is given in the next section.

\section{Discussion of the Amplification}
\subsection{Single field case}
\begin{figure}
\centering
\includegraphics[width=\textwidth]{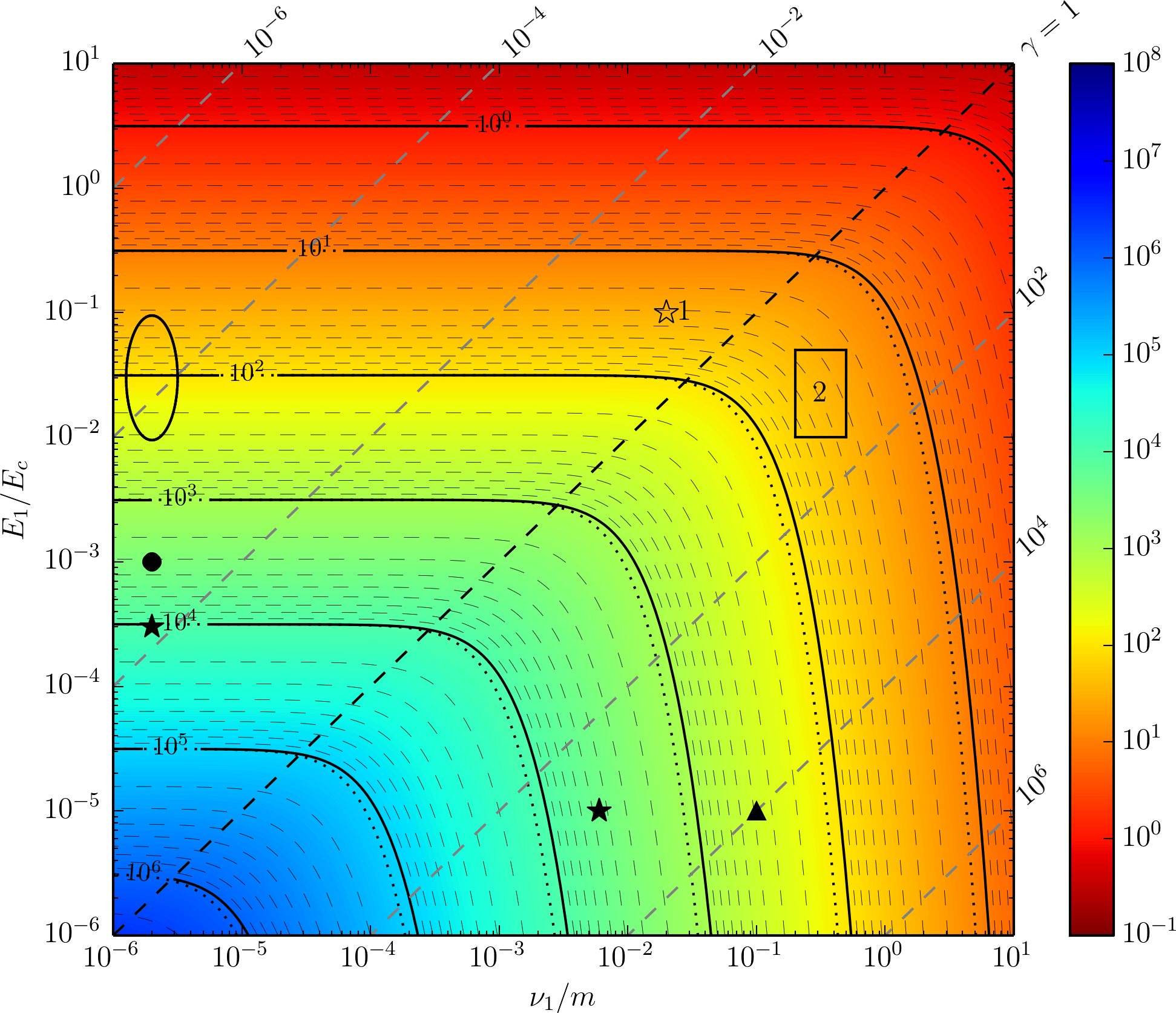}
\caption{
Contour plot of the exponential $4\frac{m}{\nu_1}G_1=4\frac{m}{\nu_1}G(p_\perp
\ll m,\gamma_1,\gamma_2\rightarrow\infty)$ of the single-field spectral envelope
function over the field-frequency ($E_1/E_c$ vs.\ $\nu_1/m$) plane (solid
curves: using~\eqref{eq:G} for $G$, dotted curves: the approximations~(\ref
{eq:G_approx},\ref{eq:g_approx})).
Note that the spectral envelope is given by $\frac{2}{9}\nu_1^2\ftt^2\env_1$
with $\env_1=\exp\left(-4\frac{m}{\nu_1}G_1\right)$.
Lines (dashed) of constant Keldysh parameters are displayed too.
Parameters from table~1 in~\cite{ringwald_pair_2001} are depicted by stars,
while the bullet is for ELI-NP~\cite{eli-np};
the ellipse is a former perspective of ELI used, e.g., in~\cite
{schutzhold_dynamically_2008, dunne_catalysis_2009}.
The triangle is for MaRIE at LANL~\cite{carlsten_marie_2011}.
Point 1 and the square 2 depict parameters considered in~\cite
{otto_lifting_2015}, where the validity of Eqs.~\eqref{eq:f_shells}-\eqref{eq:G}
can be checked by numerical integration of the quantum kinetic equations, as is
also exemplified in Fig.~\ref{fig:cut_env}.
Light grey dashed contour curves are for $2,\cdots,9$ between the solid decade
contour curves.
}
\label{fig:g_env}
\end{figure}
In the following we consider $\env=\exp\left(-4\frac{m}{\nu_1} G(p_\perp,
\gamma_1,\gamma_2,N)\right)$ as the essential part of the spectral envelope
of the phase space distribution.
The benchmark for the following discussion is the reference distribution for one
periodic field, i.e.\ $\gamma_2\rightarrow\infty$ as realized by
$E_2\rightarrow0$.
The exponential $4\frac{m}{\nu_1}G_1$ of the spectral envelope function $\env_1$
is displayed in Fig.~\ref{fig:g_env}.
The region $\gamma_1<1$ is often termed adiabatic (tunneling) region where the
residual pair density is independent of frequency, while $\gamma_1>1$ is the
anti-adiabatic (multi-photon) region which is weakly dependent on the field
strength.
Focusing on low-momentum particles, $p_\perp/m\ll1$, the spectral envelope in
the adiabatic region behaves as $\exp\bigl(-\pi(E_c/E_1)\bigr)$, while in the
anti-adiabatic region it behaves like $\exp(-4\frac{m}{\nu_1}\log4\gamma_1+4
\frac{m}{\nu_1})$ in leading order based on~\eqref{eq:G_approx}.
The contours (solid curves) in Fig.~\ref{fig:g_env} are based on~\eqref{eq:G};
they differ marginally on the approximate estimates~\eqref{eq:G_approx} (dotted
curves).
Figure~\ref{fig:g_env} provides a quantitative description of the landscape of
the dynamical Schwinger effect, first qualitatively discussed in~\cite
{blaschke_properties_2013}.
Note the huge variation over the displayed parameter range by recollecting that
the residual phase density at $p_\parallel=0$ is bounded from above by
$\frac{2}{9}\nu_1^2\ftt^2\env_1$ with $\env_1=\exp\left(-4\frac{m}{\nu_1}G_1
\right)$, i.e.\ in the blue region one meets a very strong suppression of the
pair density, guaranteeing a long lifetime of the vacuum.
Only the red region above the $\gamma_1=1$ line seems to allow for verifying
experimentally Schwinger's tunneling process in one oscillating field.
As pointed out in~\cite{otto_lifting_2015} (see also~\cite
{akal_electron-positron_2014} and further references therein), the impact of a
second field can significantly reduce the strong suppression due to large values
of $4\frac{m}{\nu_1}G$, i.e., it results in an amplification effect, as
mentioned above.

\subsection{Two field case: Amplification}
The key for an interpretation of the amplification effect by an assisting field
($\nu_2$, $E_2$) is the approximation~\eqref{eq:G_approx} for $p_\perp\ll m$:
Since $x_0(\gamma_1,\gamma_2,N)<x_0(\gamma_1,\gamma_2\rightarrow\infty)$, the
spectral envelope function $\env_1=\exp\left(-4\frac{m}{\nu_1}G_1\right)$ is
lifted by the amount $\env_{1+2}/\env_1=\exp\bigl(4\frac{m}{\nu_1}\bigl[-G(
\gamma_1,\gamma_2,N)+G_1(\gamma_1)\bigr]\bigr)\approx\exp\bigl( \pi\frac{m}
{\nu_1}\bigl[-x_0(\gamma_1,\gamma_2,N)+x_0(\gamma_1,\gamma_2\rightarrow\infty)
\bigr] \bigr)$, where $\env_{1+2}=\exp\left(-4\frac{m}{\nu_1}G(\gamma_1,
\gamma_2,N)\right)$.
Due to the usually considered range of values $\frac{m}{\nu_1}$, even a moderate
reduction of $G$ (or $x_0$) due to the assisting field, e.g.\ by $\mathcal O(50
\%)$, leads to a huge reduction of the suppression in the subcritical region
$E_1/E_c\ll1$ and $\nu_1/m\ll1$ steered either by $\frac{m}{\nu_1}\gamma_1=E_c/
E_1$ in the adiabatic region or by $\frac{m}{\nu_1}\log\gamma_1$ in the
anti-adiabatic region.
\begin{sidewaysfigure}
\centering
\includegraphics[width=\textwidth]{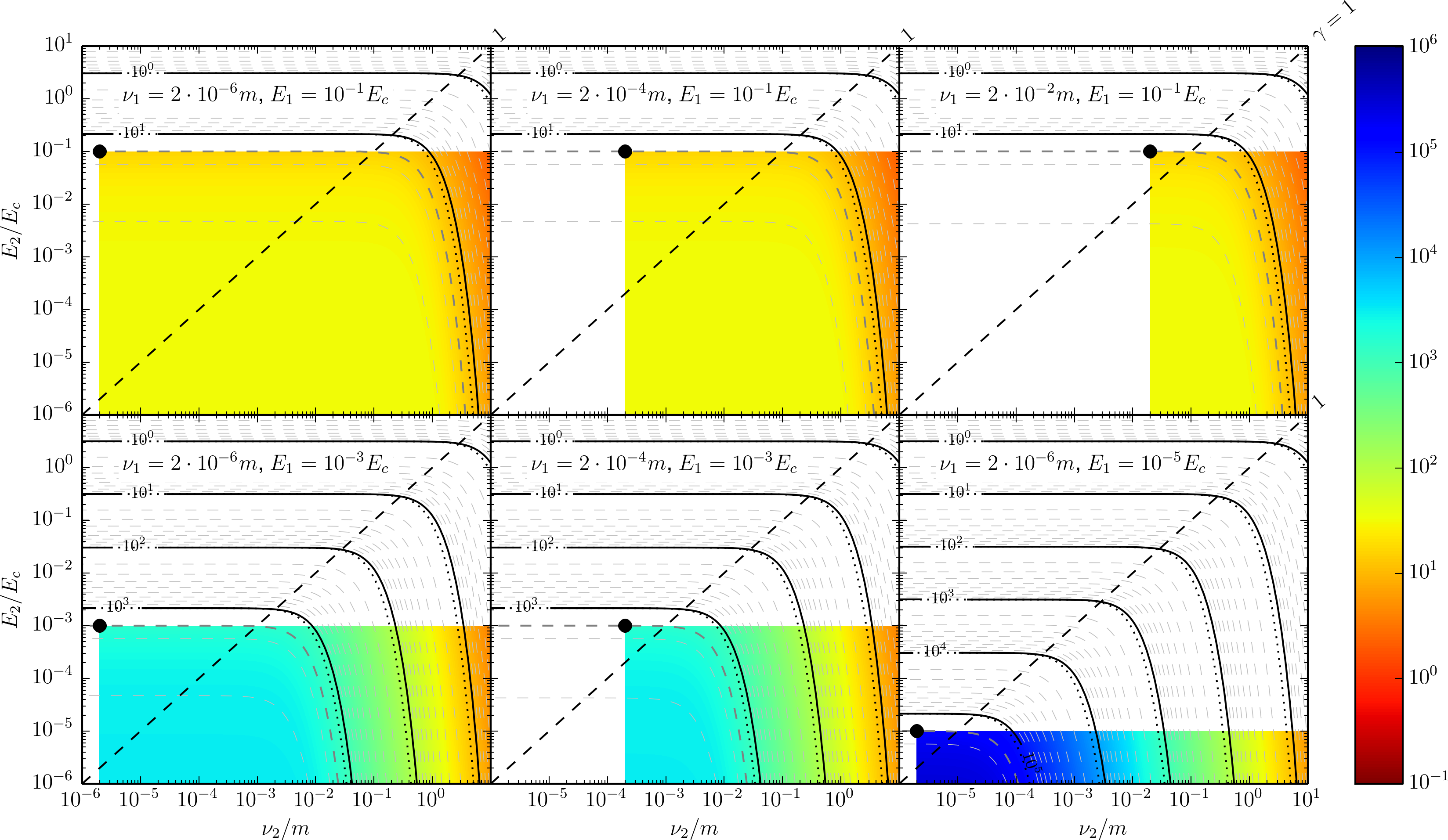}
\caption{
Contour plots of the exponential $4\frac{m}{\nu_1}G(p_\perp\ll m,\gamma_1,
\gamma_2,N)$ for six given fields $\nu_1$, $E_1$ in the adiabatic region
(positions depicted by the bullets, which are the loci of field doubling) over
the field-frequency ($E_2/E_c$ vs.\ $\nu_2/m$) plane, i.e.\ actually $4\frac{m}
{\nu_1}G(p_\perp\ll m,\nu_1,E_1,\nu_2,E_2)$.
Despite of the displayed smooth distribution, our results are strictly valid
only for $E_2<E_1$ and $\nu_2=(4n+1)\nu_1$, $n=0,1,2\cdots$.
Light grey dashed contour curves are for $2,\cdots,9$ between the solid decade
contour curves.
The heavy grey dashed curves are constructed to go through the bullets.
An amplification beyond the field doubling occurs in the colored regions right
to the heavy grey dashed curves.
}
\label{fig:env_multi}
\end{sidewaysfigure}
A few examples are exhibited in Fig.~\ref{fig:env_multi}, where we show the
exponential $4\frac{m}{\nu_1}G(p_\perp\ll m,\gamma_1,\gamma_2,N)$, entering the
spectral envelope $\frac{2}{9}\nu_1^2\ftt^2\env_{1+2}$, for a few selected
fields $E_1$, $\nu_1$ over the $E_2$-$\nu_2$ plane.
As reference one has to take the spectral envelope function from Fig.~\ref
{fig:g_env}.
The value of the envelope exponential is trivially reduced by $50\%$ in the
points $E_2=E_1$ and $\nu_2=\nu_1$ (marked by symbols in Fig.~\ref{fig:g_env}),
i.e.\ simply doubling the field strength.
This is the lowest order of the multi-field configurations considered in~\cite
{narozhny_pair_2004}. 
Further amplification occurs for enlarging $\nu_2$, as signalled by a reduction
of $4\frac{m}{\nu_1}G$.
That is, the amplification beyond the field doubling happens in the region right
to the contour curve (heavy grey dashed) going through the reference point
(bullet) $E_2=E_1$, $\nu_2=\nu_1$.
The single field result is recovered by inspection of $E_2\rightarrow0$,
i.e.\ going down from the reference point at $\nu_1$, $E_1$.
Note that our amplification estimate by formula~\eqref{eq:F_exp} requires $E_2<
E_1$ and $\nu_2=N\nu_1$ with $N=4n+1$, $n=0,1,2\cdots$, that is our result holds
true actually only at discrete values of $\nu_2$.
\begin{figure}
\centering
\includegraphics[width=0.8\textwidth]{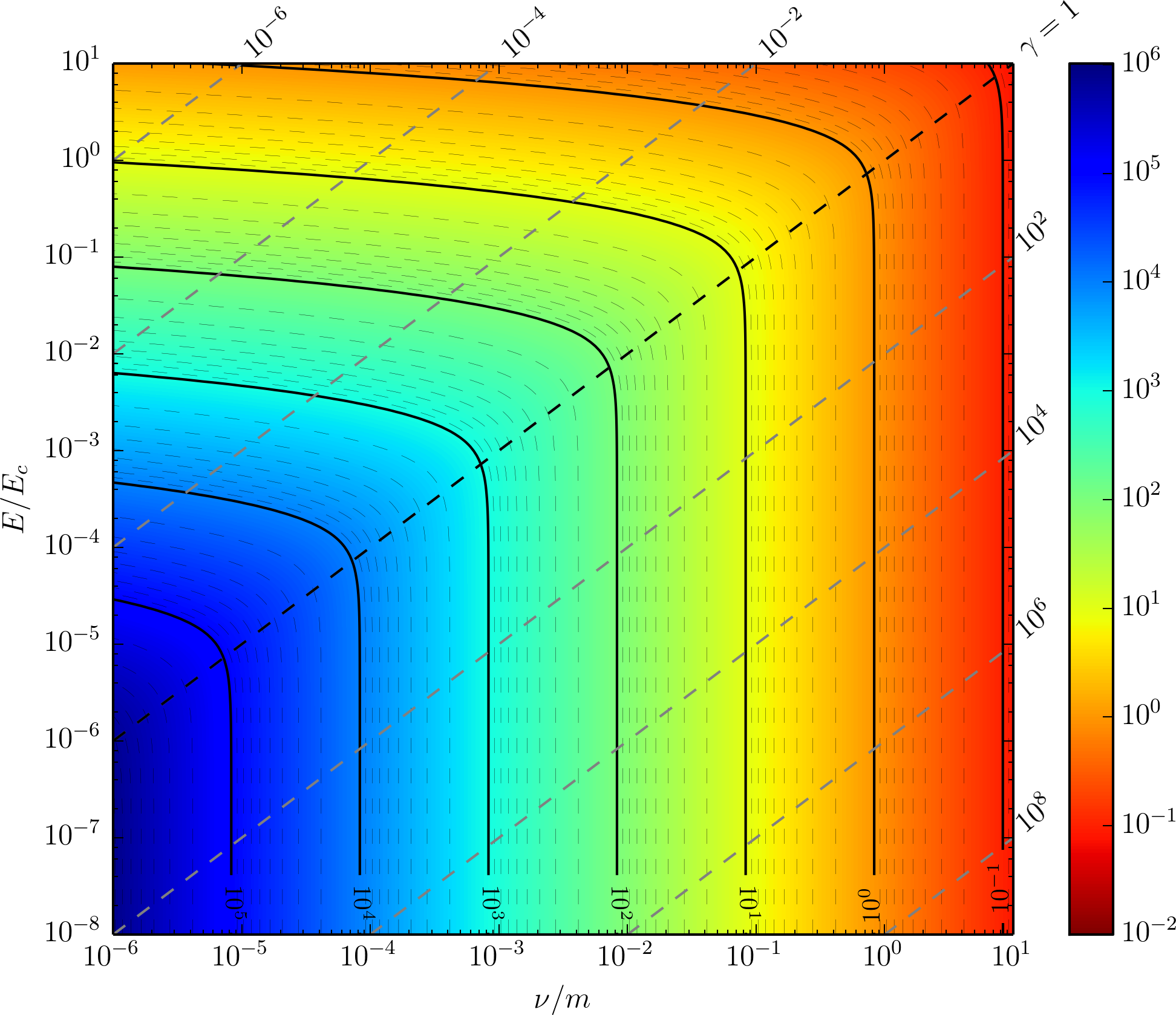}
\caption{
Contour plot of the number of main peaks in the inverval $p_\perp=0\cdots m$ at
$p_\parallel=0$ for the action of one field characterized by $\nu$ and $E$.
Light grey dashed contour curves are for $2,\cdots,9$ between the solid decade
contour curves.
Dashed lines depict constant Keldysh parameters.
}
\label{fig:state_dens}
\end{figure}

The net outcome points to the importance of the frequency $\nu_2$, while
variations of $E_2$ are subleading.
To achieve amplifications which overcome the strong suppression in the tunneling
regime one has to employ frequencies $\nu_2$ near or above critical range
$\nu\gtrsim m$, where essentially the multi-photon process sets the scale,
i.e.\ one has to compare the phase space population with the one for the
$\nu_2$, $E_2$ field alone.
This is accomplished by inspecting Fig.~\ref{fig:g_env} for the exponential of
the spectral envelope function and by correcting the pre-exponential factor
in~\eqref{eq:F_exp} by $\nu_1\rightarrow\nu_2$.
As shown in~\cite{otto_lifting_2015} for selected examples (see also
Fig.~\ref{fig:cut_env}), the action of the field ($\nu_2$, $E_2$) lifts the
phase space distribution emerging from ($\nu_1$, $E_1$) alone above the one of
($\nu_2$, $E_2$) alone.
A decisive difference is in the density of states (peaks) in the $p_\perp
(p_\parallel)$ distributions:
The number of peaks within $p_\perp=0\cdots m$ is given by $\frac{2}{\nu_1}\bigl[
\Omega(p=m,\varphi=0,\gamma_1,\gamma_2,N)-\Omega(p=0,\varphi=0,\gamma_1,\gamma_2
,N)\bigr]$, which can be much larger for the field ($\nu_1$, $E_1$) alone than
the field ($\nu_2$, $E_2$) alone, depending on the locus in the $\nu$-$E$ plane,
see Fig.~\ref{fig:state_dens}.

\section{Discussion}
The goal of our paper is to provide a qualitative understanding for the
amplification of the pair creation rate in a periodic electric field under the
resonant assistance of a second faster and weaker electric field.
In the quantum kinetic approach, such an investigation would require scans of
the four-dimensional parameter space $E_{1,2}$, $\nu_{1,2}$ for a dense set of
points in the two-dimensional $p_\perp$-$p_\parallel$ phase space.
In addition, the impact of the temporal structure (details of switching on/off
and duration of the flat top time span) should be considered.
To avoid such a cumbersome wealth of information we refrain from considering the
harsh landscape of the phase space distribution by analysing only its spectral
envelope instead.
As a consequence, we do not need to consider the details of the peak (ridge)
positions, which are modified by the assistance field.
We also rely on the previous observation~\cite{otto_lifting_2015} concerning the
dependence of the residual phase space distribution on the time structure.
This in turn is based on a harmonic analysis in the low-density
approximation~\cite{otto_lifting_2015}.

Under these conditions, the amplification, observed in several previous
investigations which are restricted to small patches in the parameter space, can
be explained as a reduction of the huge suppression provided by $\exp(-\pi(E_c/
E_1)x_0)$ in the adiabatic or by $\exp(-4(m/\nu)x_0)$ in the anti-adiabatic
regimes:
by the assistance field, the quantity $x_0$ (the zero of a simple function)
becomes smaller.
Applying these findings to typical parameters which represent the optical
laser--XFEL combination according to~\cite{ringwald_pair_2001} we find a
promising perspective for laser intensities only significantly above the present
ELI-NP plan~\cite{eli-np}.
More promising is the laser-$\gamma$ beam combination, as already pointed out
in~\cite{schutzhold_dynamically_2008, dunne_catalysis_2009} for ultra-intense
laser beams, but we see also a good discovery potential for ELI-NP~\cite{eli-np}
and even multi-PW lasers.
The potentially dangerous $\gamma$ conversion processes might be rejected by the
phase space distribution of the residual pairs:
The amplified tunneling production displays the distinct peak (ridge) structure
dictated by the frequency $\nu_1$.

To arrive at such a simple picture we made two restrictions:
(i) $N=4n+1$, $n=0,1,2,\cdots$ was selected since the main pole in the complex
time plane is then shifted in purely imaginary direction towards the real axis
under the action of the assistance field (see Appendix), thus allowing for an
easy access to the spectral envelope function and
(ii) $p_\parallel=0$ was chosen as it was shown in~\cite{otto_lifting_2015} to
provide a proper orientation.
The avenues for next generalizations are obvious:
extensions to arbitrary integer $N$ (and further also real $N$ which go beyond
the resonant type patterns) as well as further poles and estimates of the phase
space-integrated residual distribution to arrive at a density.
Still, the model is fairly simple and needs further considerations of important
effects, such as spatial inhomogenities which are, for example, elaborated
in~\cite{dunne_worldline_2005,ruf_pair_2009,schneider_dynamically_2014}.

\section{Summary}
In summary we provide a comprehensive tool to access the amplification of the
dynamical Schwinger process by a resonantly assisting second field, both acting
for a finite duration.
A promising perspective is seen in the combination of ultra-intense optical
laser beams with $\gamma$ beams, while the optical laser-XFEL combination also
allows for huge amplification effects, however, probably not sufficient to
enable an experimental verification.

\section*{Acknowledgments}
T.~E.~Cowan and R.~Sauerbrey are gratefully acknowledged for a fruitful
collaboration within the HIBEF project at European XFEL.
The authors thank R.~Alkofer, H.~Gies, S.~S.~Schmidt, and R.~Sch\"utzhold for
inspiring discussions.
A.~Ringwald pointed out the paramount importance of~\cite{popov_resonant_1973,
ringwald_pair_2001} for the topic.
The work of D.~Blaschke was supported in part by the Polish Ministry of Science
and Higher Education (MNiSW) under grant no.\ 1009/S/IFT/14.
Two of the authors (DS and BK) acknowledge the hospitality of KITP\slash Santa
Barbara, where a part of the investigation has been performed during the KITP
program ``Frontiers of Intense Laser Physics''.
During that period DS was supported in part by the National Science Foundation
under grant no.\ NSF PHY11-25915.

\section*{Appendix: Derivation of Eqs.~\eqref{eq:f_shells}-\eqref{eq:G}}
In the low-density approximation of the quantum kinetic equation~\cite
{schmidt_quantum_1998,mamayev_mostepanenko}, the electron-positron pair density
$f$ is given by $f(\vec p,t)=\frac{1}{2}\left|I(\vec p,t)\right|^2$ with
\begin{align}
I(\vec p,t) &= \lint_0^t\!\frac{eE(t')\eperp}{\omega(\vec p, t')^2} \eh{2i
\Theta(\vec p,t')}\diff t'\:,
\end{align}
where we introduced $\eperp=\sqrt{m^2+p_\perp^2}$, $\omega(\vec p,t)=\sqrt{m^2
+p_\perp^2+\bigl(p_\parallel-eA(t)\bigr)^2}$, $\Theta(\vec p,t)=\lint_0^t\!
\diff t'\omega(\vec p,t')$.
The vector potential $A(t)$ and hence the electric field $E(t)=-\dot A(t)$ are
assumed to be periodic in time, e.g.\ $A=(E_1/\nu_1)\cos(\nu_1t)+(E_2/(N\nu_1))
\cos(N\nu_1t)$ over a finite time span $\ftt$, much longer than the
switching-on/off intervals.
Accordingly one can split $\Theta$ in a linearly growing and a periodic part,
$\Theta(\vec p,t)=\Omega(\vec p)t+P(\vec p,t)$, to arrive at $I(\vec p,t)=
\lint_0^t\!\!\diff t' F(\vec p,t')\eh{2i\Omega(\vec p)t'}$ with $F(\vec p,t)=
\frac{eE(t)\eperp}{\omega(\vec p, t)^2}\eh{2iP(\vec p,t)}$.
The function $F$ is time periodic and is decomposed into Fourier modes,
$F(\vec p,t)=\sum_\ell F_\ell(\vec p)\eh{-i\ell\nu t}$, so that $I$ turns into
\begin{align}
I(\vec p,t) = \sum_\ell iF_\ell(\vec p) \frac{\eh{-i\bigl(\ell\nu-2\Omega(
\vec p)\bigr)t}-1}{\ell\nu-2\Omega(\vec p)}\:.
\end{align}
The quantity $I$ becomes large whenever $2\Omega(\vec p)-\ell\nu$ is small,
yielding~\eqref{eq:shell_cond} and~\eqref{eq:Omega}.
Employing polar coordinates one defines the ridge radius $p_\ell(\varphi)$ by
$2\Omega\bigl(p_\ell(\varphi),\varphi\bigr)-\ell\nu=0$.
Since $\Omega(p,\varphi)$ as a function of $p$ is strictly monotonous, this
resonance equation has solutions only for $\ell\ge\ell_\text{min}=\left\lceil
\frac{2\Omega(0,\varphi)}{\nu}\right\rceil$.
Expanding every summand in $I$ around its corresponding peak position at
$p_\ell$ to first order, and keeping $F_\ell$ fixed, we arrive at
\begin{align}
I(\vec p, t) = \sum_{\ell=\ell_\text{min}}iF_\ell\bigl(p_\ell(\varphi),
\varphi\bigr)\frac{\exp\left[2i\Omega'\bigl(p_\ell(\varphi),\varphi\bigr)
\bigl(p-p_\ell(\varphi)\bigr)t\right]-1}
{2i\Omega'\bigl(p_\ell(\varphi),\varphi\bigr)\bigl(p-p_\ell(\varphi)\bigr)}\:;
\label{eq:I_expanded}
\end{align}
terms with $\ell<\ell_\text{min}$ do not contribute and are thus dropped.
The prime denotes the derivative w.r.t.\ $p$, i.e.\ $\Omega'(p,\varphi)=\partial
\Omega(p,\varphi)/\partial p$.
To arrive at $f$ one must take the squared modulus of~\eqref{eq:I_expanded}.
Terms mixing different $\ell$s in this square will go to zero upon switching
off, as one can argue from a slowly varying envelope approximation.
So only the non-mixing terms survive for $t>\ftt$, yielding in particular the
$\sin^2x/x^2$ part of Eq.~\eqref{eq:f_shells}, which contains ridge positions
and ridge widths as anticipated in~\cite{otto_lifting_2015}.
For larger times $\ftt$ and small momenta, the Popov formula~\cite
{popov_resonant_1973,ringwald_pair_2001}) is recovered for the special case
$E_2=0$.
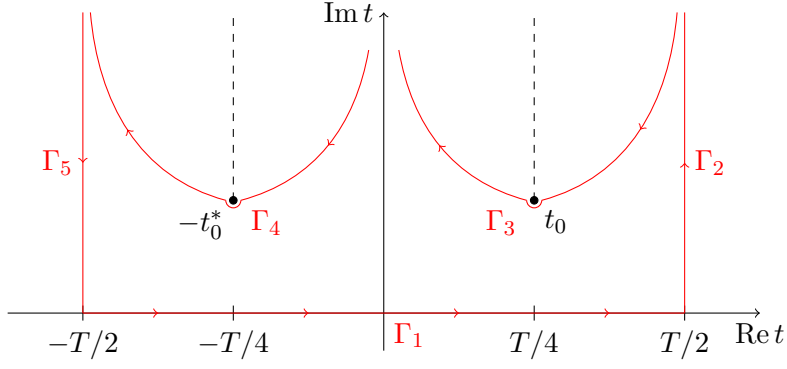
\begin{figure}
\centering
\begin{tikzpicture}
\draw[->] (-5,0) -- (5,0) node[below] {$\Re t$};
\draw[->] (0, -0.5) -- (0,4) node[left] {$\Im t$};
\draw[-] (-4, 0.1) -- (-4, -0.1) node[below] {$-T/2$};
\draw[-] (-2, 0.1) -- (-2, -0.1) node[below] {$-T/4$};
\draw[-] (2, 0.1) -- (2, -0.1) node[below] {$T/4$};
\draw[-] (4, 0.1) -- (4, -0.1) node[below] {$T/2$};

\draw[dashed] (2, 1.5) -- (2, 4);
\draw[dashed] (-2, 1.5) -- (-2, 4);
\filldraw (2, 1.5) node[below right] {$t_0$} circle(0.05);
\filldraw (-2, 1.5) node[below left] {$-t_0^*$} circle(0.05);

\begin{scope}[decoration={
    markings,
    mark=at position 0.5 with {\arrow{>}}}
    ]
\draw[red, postaction={decorate}] (-4, 4) -- (-4, 0);
\node[red, left] at (-4, 2) {$\Gamma_5$};
\draw[red, postaction={decorate}] (-4, 0) -- (-2, 0);
\draw[red, postaction={decorate}] (-2, 0) -- (0, 0) node[below right] {$\Gamma_1$};
\draw[red, postaction={decorate}] (0, 0) -- (2, 0);
\draw[red, postaction={decorate}] (2, 0) -- (4, 0);
\draw[red, postaction={decorate}] (4, 0) -- (4, 4);
\node[red, right] at (4, 2) {$\Gamma_2$};

\draw[red, postaction={decorate}] (3.9, 4) to[out=265, in=15] (2.1, 1.5);
\draw[red, postaction={decorate}] (1.9, 1.5) node[below left] {$\Gamma_3$} to[out=165, in=280] (0.2, 3.5);
\draw[red] (1.9, 1.5) arc[radius = 0.1, start angle= 180, end angle= 360];

\draw[red, postaction={decorate}] (-0.2, 3.5) to[out=260, in=15] (-1.9, 1.5) node[below right] {$\Gamma_4$};
\draw[red, postaction={decorate}] (-2.1, 1.5) to[out=165, in=275] (-3.9, 4);
\draw[red] (-2.1, 1.5) arc[radius = 0.1, start angle= 180, end angle= 360];
\end{scope}
\end{tikzpicture}
\caption{
The integration path $\Gamma_1$ from $-T/2$ to $+T/2$ along the real axis
in~\eqref{eq:fourier_coeff} is substituted by the sequence $\Gamma_2+\Gamma_3+
\Gamma_4+\Gamma_5$ in the upper half of the complex $t$ plane.
Dots depict the zeros of $\omega(t)^2$, where branch cuts (dashed lines) emerge.
}
\label{fig:cont}
\end{figure}

What remains is a formula for the Fourier coefficients $F_\ell$, which determine
the ridge heights.
This can be achieved by deforming the integration contour in the complex time
plane and using the method of steepest descent, similar to~\cite
{brezin_pair_1970}.
For the single field case, the Fourier coefficients are given by
\begin{align}
F_\ell\bigl(p_\ell(\varphi),\varphi\bigl) = \frac{1}{T}\lint_{-T/2}^{T/2}\!
\frac{eE(t)\eperp}{\omega\bigl(p_\ell(\varphi),\varphi, t\bigl)^2}
\eh{2i\Theta(p_\ell(\varphi),\varphi,t)}\diff t\:.
\label{eq:fourier_coeff}
\end{align}
The function $\omega(t)^2$ has four zeros of first order in the strip $-T/2<\Re
 t<T/2$ in the complex time plane at $t_0$, $t_0^*$, $-t_0$ and $-t_0^*$.
We deform the integration contour to the sequence $\Gamma_2+\Gamma_3+\Gamma_4+
\Gamma_5$ shown in Fig.~\ref{fig:cont}.
The contributions from $\Gamma_2$ and $\Gamma_5$ cancel due to the periodicity
of the integrand.
The contribution from $\Gamma_4$ is the negative of the complex conjugate of
that from $\Gamma_3$, in symbolic notation $\int_{\Gamma_1}=-\int_{\Gamma_3}-
\int_{\Gamma_4}=-2i\Im\int_{\Gamma_3}$.
To evaluate the latter integral approximately, we note that the exponent in the
integrand is stationary at $t_0$.
One chooses $\Gamma_3$ such that $\Im\Theta$ grows rapidly on $\Gamma_3$
away from $t_0$, so that $\eh{2i\Im\Theta}$ quickly goes to zero (method of
steepest descent).
Then only contributions to the integral close to $t_0$ matter, thus enabling an
expansion $\omega(t)^2 = 2i\eperp eE(t_0)(t-t_0)+\mathcal O\bigl((t-t_0)^2
\bigr)$ yielding (we suppress the momentum arguments)
\begin{align}
\frac{eE(t)\eperp}{\omega(t)^2} = \frac{1}{2i}\frac{1}{t-t_0}+\mathcal O(1)\:.
\label{eq:frac_expansion}
\end{align}
The expansion of the exponent uses $\Theta(t)=\int_0^t\!\diff t'\omega(t')=
\int_0^{t_0}\!\diff t'\omega(t')+\int_{t_0}^{t}\!\diff t'\omega(t')=\Theta(t_0)+
\int_{t_0}^{t}\!\diff t'\sqrt{2eE(t_0)\eperp}\sqrt{i(t'-t_0)}+\mathcal O\bigl(
(t-t_0)^\frac{5}{2}\bigr)$ with the result
\begin{align}
\Theta(t) = \Theta(t_0) - \frac{2}{3}i\sqrt{2eE(t_0)\eperp}\bigl(i(t-t_0)
\bigr)^\frac{3}{2} + \mathcal O\bigl((t-t_0)^\frac{5}{2}\bigr)\:.
\label{eq:exp_expansion}
\end{align}
Inserting~\eqref{eq:frac_expansion} and~\eqref{eq:exp_expansion} into~\eqref
{eq:fourier_coeff} yields our desired result
\begin{align}
F_\ell = -\frac{1}{T}i\Im\lint_{\Gamma_3}\!\!\diff t\,\eh{2i\Theta(t_0)}
\frac{\eh{\frac{4}{3}\sqrt{2eE(t_0)\eperp}(i(t-t_0))^\frac{3}{2}}}
{i(t-t_0)} = \frac{4\pi i}{3T}\Im\eh{2i\Theta(t_0)} = \frac{2}{3}i\nu\Im\eh{2i
\Theta(t_0)}\:.
\end{align}
To generalize to two fields we need to sum over all complex zeros $t_i$ of
$\omega^2$ with $0\le\Re t_i<T/2$, $T=2\pi/\nu_1$, and $\Im t_i>0$, leading to
\begin{figure}
\centering
\begin{tikzpicture}
\draw[->] (-5,0) -- (5,0) node[below] {$\Re t$};
\draw[->] (0, -0.5) -- (0,4) node[left] {$\Im t$};
\draw[-] (-4, 0.1) -- (-4, -0.1) node[below] {$-T/2$};
\draw[-] (-2, 0.1) -- (-2, -0.1) node[below] {$-T/4$};
\draw[-] (2, 0.1) -- (2, -0.1) node[below] {$T/4$};
\draw[-] (4, 0.1) -- (4, -0.1) node[below] {$T/2$};
\draw[-] (-0.1, 3) -- (0.1, 3) node[right] {$T/10$};

\draw[mark=*] plot coordinates {(2,10*3*0.19869/3.141592653)};
\draw[mark=*] plot coordinates {(-2,10*3*0.19869/3.141592653)};

\draw[->] (2,10*3*0.18/3.141592653) -- (2,10*3*0.11/3.141592653);
\draw[->] (-2,10*3*0.18/3.141592653) -- (-2,10*3*0.11/3.141592653);

\foreach \x/\y in {0.118687/0.185214, 3.02291/0.185214, 2.7702/0.182433, 0.371393/0.182433, 2.51792/0.17675, 0.623669/0.17675, 2.26631/0.167189, 0.875288/0.167189, 2.01635/0.151386, 1.12524/0.151386, 1.77416/0.123266, 1.36744/0.123266, 1.5708/0.0945343}
{
\draw[mark=square*, red] plot coordinates {(4*\x/3.141592653,3*10*\y/3.141592653)};
\draw[mark=square*, red] plot coordinates {(-4*\x/3.141592653,3*10*\y/3.141592653)};
}
\end{tikzpicture}
\caption{
The distribution of the zeros of $\omega^2$ in the upper half plane for field
``1+2'' (red squares) for the example of $N=25$.
The two zeros for field ``1'' are shown as black dots, which get shifted down
(arrows) upon the impact of field ``2''.
}
\label{fig:zeros}
\end{figure}
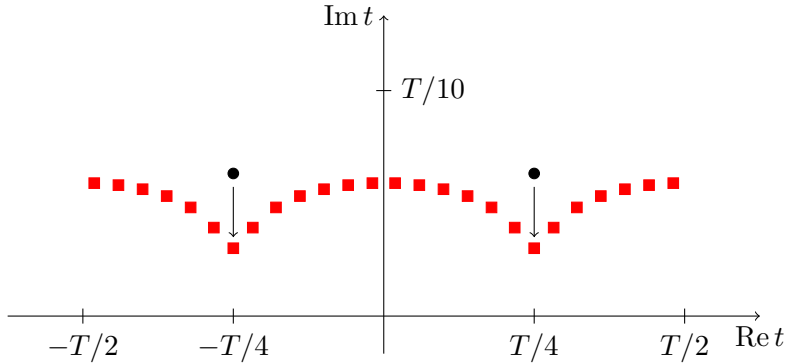
\begin{align}
F_\ell\bigl(p_\ell(\varphi),\varphi\bigr) \approx \frac{2}{3}i\nu \sum_i\Im\eh
{2i\Theta(p_\ell(\varphi),\varphi,t_i)}\:.
\label{eq:F_sum}
\end{align}
It is instructive to take $\varphi=0$, that is we consider the phase space
distribution for $p_\parallel=0$.
For the case $N=4n+1$, integer $n$, a sequence of $N+1$ zeros appears at $\Im t>
0$, $-T/2<\Re t<T/2$ (see Fig.~\ref{fig:zeros}, red squares).
In particular, the zeros at $\pm T/4+ix_0/\nu_1$ get shifted down, since $x_0$
is smaller in the two-field case than in the single-field case.
The sum in~\eqref{eq:F_sum} is dominated by the contribution with the smallest
imaginary part, which is the zero $t_0=T/4+ix_0/\nu_1$.
Keeping this leading term and using furthermore $\Theta(T) = 4\Theta(T/4)$
yields
\begin{align}
\Theta(t_0) = \frac{\ell\pi}{4} + i\frac{m}{\nu}\lint_0^{x_0}\!\!\diff x
\sqrt{1+\frac{p_\perp^2}{m^2} - \left(\frac{1}{\gamma_1}\sinh x
+\frac{1}{\gamma_2}\sinh Nx\right)^2}\:,
\end{align}
finally leading to~\eqref{eq:F_exp} and~\eqref{eq:G}.
The amplification effect by a second assistance field is thus rooted in a shift
of the leading-order pole towards the real axis in the complex $t$ plane.
The condition $E_2\le E_1$ ensures the pattern exhibited in Fig.~\ref{fig:zeros}
and $N=4n+1$ is specific for the shift of the single field zeros (black dots in
Fig.~\ref{fig:zeros}) parallel to the imaginary $t$ axis.
\bibliographystyle{custom}
\bibliography{lit}
\end{document}